\documentclass[twocolumn,preprintnumbers,superscriptaddress,nofootinbib,aps,prd,floatfix]{revtex4}

\pdfoutput=1

\usepackage{amsmath,amssymb}
\usepackage{graphicx} 
\usepackage{epstopdf} 
\usepackage{slashed}
\usepackage{color}
\usepackage{multirow}
\usepackage{footmisc}
\usepackage{hyperref}
\usepackage{subfigure}
\usepackage{snapshot}
\usepackage{hepunits}

\usepackage{array}
\newcolumntype{L}[1]{>{\raggedright\let\newline\\\arraybackslash\hspace{0pt}}m{#1}}
\newcolumntype{C}[1]{>{\centering\let\newline\\\arraybackslash\hspace{0pt}}m{#1}}
\newcolumntype{R}[1]{>{\raggedleft\let\newline\\\arraybackslash\hspace{0pt}}m{#1}}

\DefineFNsymbols*{lamport}{\dagger\ddagger\S\P\|{**}{\dagger\dagger}{\ddagger\ddagger}}

\renewcommand{\thefootnote}{\fnsymbol{footnote}}

\renewcommand{\d}{{\mathrm{d}}}
\newcommand{\be}{\begin{eqnarray*}}
\newcommand{\ee}{\end{eqnarray*}}

\newcommand{\bee}{\begin{eqnarray}}
\newcommand{\eee}{\end{eqnarray}}
\newcommand{\beeq}{\begin{equation}}
\newcommand{\eeeq}{\end{equation}}

\renewcommand{\vec}{\bf}

\begin{document}

\title{Triple Higgs boson production at a 100~TeV proton-proton collider}
%
%

\begin{abstract}
We consider triple Higgs boson production at a future 100~TeV proton-proton
collider. We perform a survey of viable final states and compare and
contrast triple production to Higgs boson pair production. Focussing on the $hhh \rightarrow (b\bar{b}) (b\bar{b})
(\gamma \gamma)$ final state, we construct a baseline analysis for the Standard Model scenario and simple deformations, demonstrating that the process merits investigation in the high-luminosity phase of the future collider as a new probe of the self-coupling sector of the Higgs boson. 
\end{abstract}

\author{Andreas Papaefstathiou} 
\email{apapaefs@cern.ch}
\affiliation{PH Department, TH Unit, CERN,\\ CH-1211 Geneva 23, Switzerland}
\author{Kazuki Sakurai} 
\email{kazuki.sakurai@kcl.ac.uk}
\affiliation{Department of Physics, Theoretical Particle Physics \&
  Cosmology,\\ King’s College London, United
Kingdom}

\preprint{CERN-PH-TH-2015-205, KCL-PH-TH/2015-36, LCTS/2015-25, MCnet-15-21}

\maketitle

\section{Multi-Higgs boson production at hadron colliders}
\label{sec:intro}
With the exception of very few interactions, most of the terms that
comprise the Standard Model (SM) Lagrangian have been measured or
constrained, their strengths found to be suggestively close to the
expected ones. An
important category of interactions \textit{not} directly observed are those
of the the Higgs boson with itself. The so-called `self-couplings' and their
energy dependence are crucial in determining the stability of the
vacuum. Current observations suggest that our Universe may be sitting
at a metastable false vacuum~\cite{EliasMiro:2011aa, Degrassi:2012ry,
  Buttazzo:2013uya, Branchina:2013jra, Branchina:2014usa,
  Bezrukov:2014ipa, Bednyakov:2015sca, Branchina:2015nda} and measurements of these couplings will illuminate this fact further. 

At colliders, these terms,
i.e. those proportional to $h^n$, $h$ being the Higgs boson scalar field, can be directly probed through the
simultaneous production of $(n-1)$ Higgs bosons. Unfortunately, the production
rates for processes with $n\geq 3$, i.e. more than one Higgs boson, are small, mainly due to the relatively large invariant
mass of the final state system. In particular, at the Large Hadron
Collider (LHC) with 14~TeV proton-proton
centre-of-mass energy, gluon-fusion Higgs boson pair production is expected to have
a cross section of $\sim$\unit{40}{\femtobarn}~\cite{Glover:1987nx,
    Dawson:1998py, Djouadi:1999rca, Plehn:1996wb, deFlorian:2013uza,
    Grigo:2013rya, Maltoni:2014eza, deFlorian:2015moa, Grigo:2015dia}, whereas
triple production is expected to have a rather dwarfish rate, with a cross section of 
$\mathcal{O}$(\unit{0.1}{\femtobarn})~\cite{Maltoni:2014eza}. Hence, even though there is
optimism that Higgs boson pair production will provide important
information and constraints through LHC
measurements~\cite{Baur:2002qd, Baur:2003gp, Contino:2010mh,
  Grober:2010yv, Dolan:2012rv, Papaefstathiou:2012qe, Dolan:2012ac,
  Contino:2012xk, Gillioz:2012se, Kribs:2012kz, Dawson:2012mk, Baglio:2012np, Barr:2013tda,
  Dolan:2013rja, Maierhofer:2013sha, No:2013wsa, Nishiwaki:2013cma,
  Liu:2013woa, Enkhbat:2013oba, Heng:2013cya, Nhung:2013lpa,
  Galloway:2013dma, Ellwanger:2013ova, Han:2013sga,McCullough:2013rea,
  Gupta:2013zza, Killick:2013mya, Choi:2013qra, Cao:2013si,
  Craig:2013hca, Goertz:2013kp, Goertz:2013eka, Goertz:2014qta,
  Englert:2014uqa,Liu:2014rva, Chen:2014xwa, Frederix:2014hta,
  Baglio:2014nea, Hespel:2014sla, Bhattacherjee:2014bca, Liu:2014rba, Wardrope:2014kya,
  Cao:2014kya, Li:2015yia, Martin-Lozano:2015dja, vanBeekveld:2015tka, Azatov:2015oxa, Liu:2015aka, Etesami:2015caa, Kang:2015nga, Dawson:2015oha, Grober:2015cwa, Lu:2015jza, He:2015spf, Dolan:2015zja, Dall'Osso:2015aia, Batell:2015koa, Dawson:2015haa}, any
direct measurement of SM-like triple Higgs boson production will be essentially impossible at the LHC,
even at the end of the high-luminosity phase
(HL-LHC)~\cite{Plehn:2005nk, Binoth:2006ym}. However,
with a significant increase in the collision energy, a Future Circular
hadron-hadron Collider (FCC-hh), colliding protons at 100 TeV, stands
a good chance at observing and constraining the self-coupling of the
Higgs bosons through Higgs boson pair production~\cite{Barr:2014sga, Kotwal:2015rba, Azatov:2015oxa, He:2015spf, Papaefstathiou:2015iba}, the cross section rising to
$\sim$\unit{1.6}{\picobarn}~\cite{deFlorian:2013jea}. Additionally, at the FCC-hh one
may also get the chance to observe three on-shell Higgs bosons being
produced, since the total cross section rises to
$\sim$\unit{5}{\femtobarn}~\cite{Maltoni:2014eza}. The
evaluation of this possibility is the main object of the present
article. 

Concretely, the part of the Higgs boson potential which includes the self-interactions, may
be written as:
\begin{equation}\label{eq:indep}
\mathcal{V}_\mathrm{self} =  \frac{ m_h^2 } { 2 v } \left( 1 + c_3 \right) h^3 +
\frac{ m_h^2 } { 8 v^2 }  \left( 1 + d_4 \right) h^4 \;,
\end{equation}
where $v \simeq 246$~GeV is the vacuum expectation value (vev), $m_h \simeq 125$~GeV is the measured Higgs boson mass and $c_3$ and $d_4$ parametrize possible deviations
from the standard model expectation (i.e. the SM is recovered for $c_3
= d_4 = 0$). 

\begin{figure*}[!htp]
  \centering
   \subfigure[]{\includegraphics[width=0.23\textwidth]{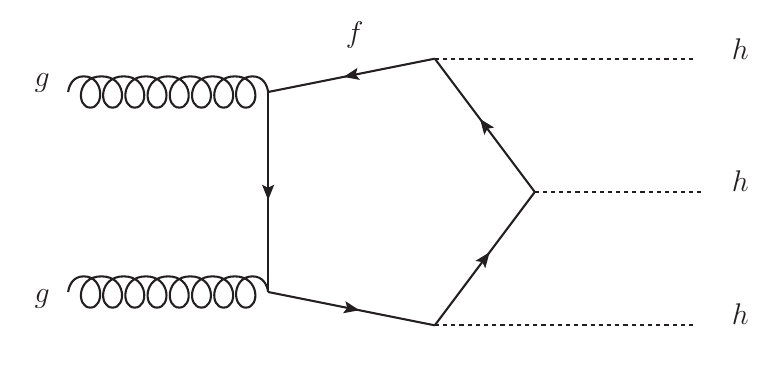}}
  \hfill
   \subfigure[]{\includegraphics[width=0.23\textwidth]{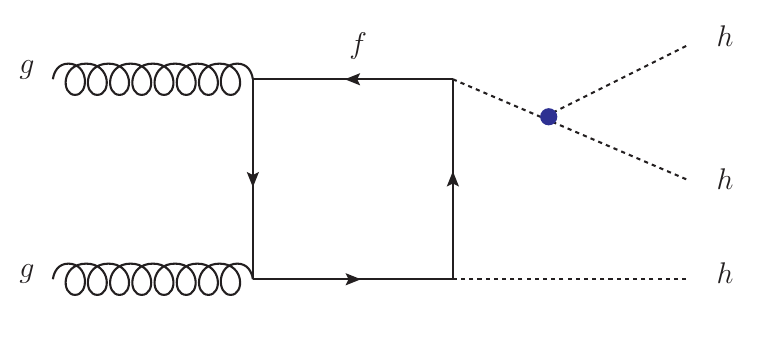}}
  \hfill
   \subfigure[]{\includegraphics[width=0.23\textwidth]{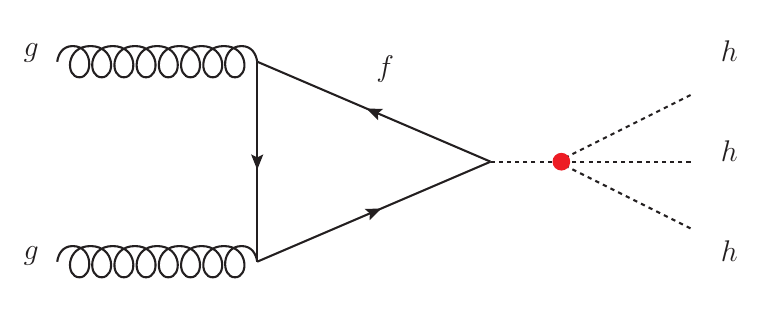}}
  \hfill
   \subfigure[]{\includegraphics[width=0.23\textwidth]{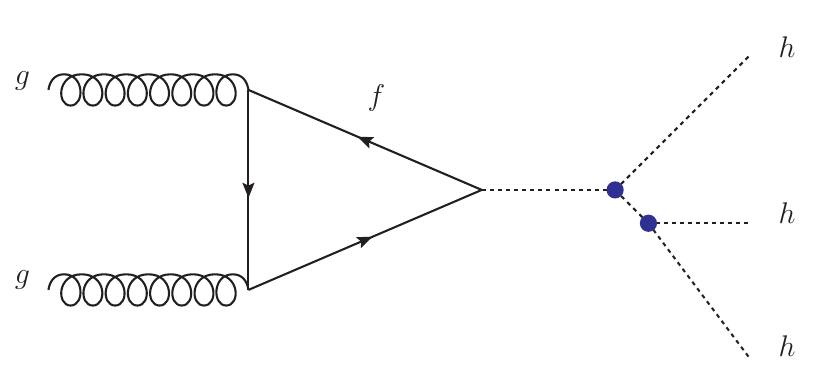}}
  \caption{\label{fig:feyndiags} Example Feynman diagrams contributing
  to Higgs boson triple production via gluon fusion in the Standard Model. The vertices
  highlighted with a blobs indicate either triple (blue) or quartic (red) self-coupling contributions.}
\end{figure*}

\begin{figure}[!htb]
  \centering
 \includegraphics[width=0.49\textwidth]{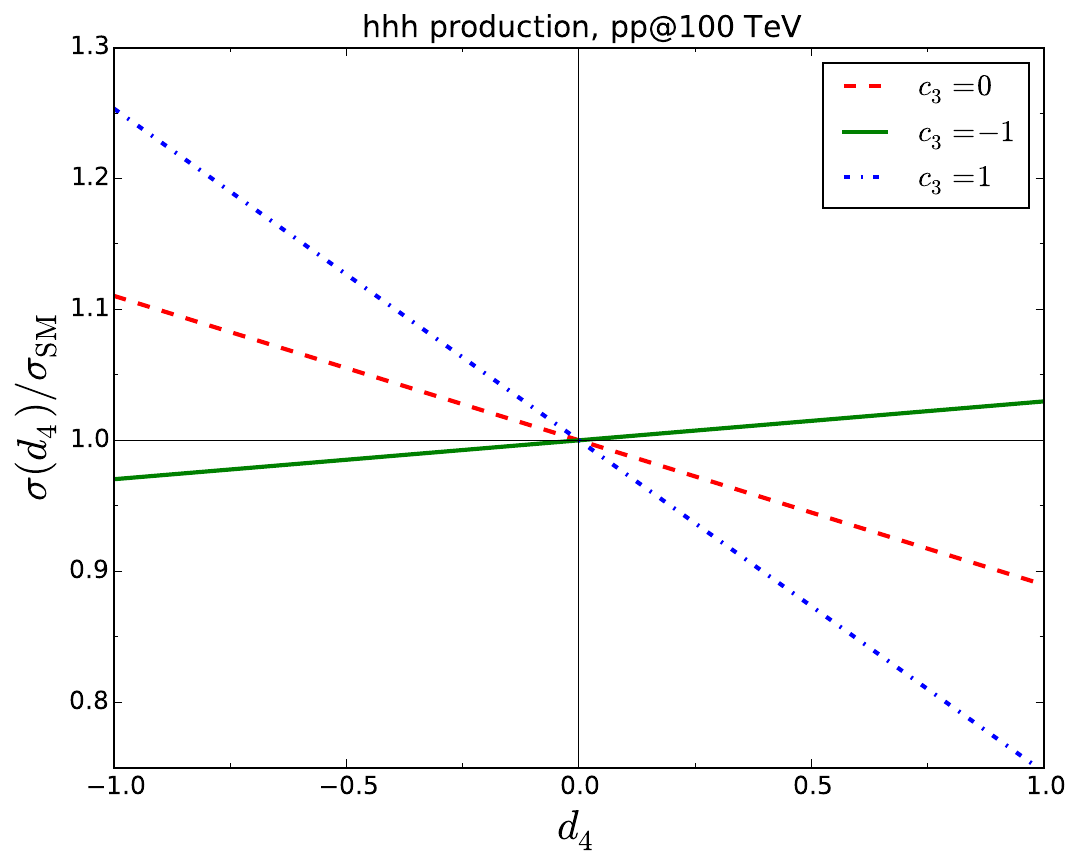}
  \caption{\label{fig:c3d4} Total cross section ratio normalised to
    the Standard Model values for gluon-fusion-initiated triple Higgs production at
    100~TeV obtained by varying the $c_3$ and $d_4$ parameters
    independently (see Eq.~\ref{eq:indep}). The Higgs boson mass was fixed to $m_h = 125$~GeV. The SM
    cross section at leading order is $\sim 2.88$~fb. The
\texttt{NNPDF23\_nlo\_as\_0119} parton density function set was used. }
\end{figure}

Figure~\ref{fig:feyndiags} shows some of the Feynman diagrams
contributing to triple Higgs boson production. It is clear that the production cross section depends on both $c_3$ and $d_4$ parameters. This should be contrasted to double Higgs boson production, which does not depend on $d_4$. In Ref.~\cite{Plehn:2005nk} the dependence of the triple Higgs boson
cross section on the parameters $c_3$ and $d_4$ was investigated at 14~TeV and
200~TeV proton-proton colliders for a Higgs boson mass $m_h = 120$~GeV. We produce an equivalent result for proton-proton
collisions at 100~TeV, for $m_h = 125$~GeV, shown in
Fig.~\ref{fig:c3d4}. The conclusions are similar to
those drawn in~\cite{Plehn:2005nk}: the cross section dependence on
$d_4$ is mild, the deviations due to $d_4 = \pm 1$ being at most $\pm
20\%$ for $c_3 = 1$. Hence modifications of the $d_4$ coefficient
itself will be very challenging
to probe. This is also demonstrated in the contour plot of
Fig.~\ref{fig:xsecs}(a), which shows the cross section normalised to
the SM value, on the $c_3-d_4$
parameter space. On this plane, one can observe that the dependence along $d_4$ is
much weaker than that along $c_3$. 

In terms of constraining $c_3$, triple Higgs boson production cannot
be superior to double Higgs boson production due to its small
production cross section. On the other hand, triple production would
be the best process to constrain $d_4$, although, as we will demonstrate, even the FCC-hh with
30~ab$^{-1}$ of integrated luminosity can only provide $\mathcal{O}(1)$ constraints on $d_4$,
because its dependency of the cross section is very modest. However,
observing the triple Higgs boson production process is an interesting task
in its own right, and as will be seen, indeed challenging at the FCC-hh. The goal of this article is to provide a first baseline
study of Standard Model-like triple Higgs boson production via gluon
fusion (ggF), at a future 100 TeV proton-proton collider.
Furthermore, we investigate triple Higgs production in two scenarios
where it is affected by new physics: (i) in the SM augmented by a single higher-dimensional operator
in an effective field theory approach and (ii) the generic case on the $(c_3 - d_4)$-plane. 

The article is organised as follows: in Section~\ref{sec:sensitivity} we investigate an explicit scenario that contains a single higher-dimensional operator. In Section~\ref{sec:fs} we list, for future reference, the
final states that could be interesting in the study of 
Higgs boson triple production. The Monte Carlo event generation, simulation of $b$-jet and photon tagging are described in
Section~\ref{sec:sim}. Differential distributions at parton level for
triple Higgs boson production at 100~TeV, compared to those of Higgs
boson pair production and the analysis of the channel $(b\bar{b})
(b\bar{b}) (\gamma \gamma)$ is described in
Section~\ref{sec:anal}. We use this analysis to provide constraints in two scenarios. Finally, we provide discussion and conclusions
in Section~\ref{sec:conc}.

\subsection{The self-coupling in $D=6$ EFT}
\label{sec:sensitivity}
In the framework of the dimension-6 operator extension to the Standard
Model ($D=6$ EFT), one can compare the sensitivity of multi-Higgs production to
variations of the operator Wilson
coefficients~\cite{Goertz:2014qta}. Here we consider, as an
illustrative example, a simplified mode with the assumption that the effect of all coefficients apart from a single one,
originating from an operator of the form $\mathcal{O}_6 \sim |H|^6$, where $H$ is the
Higgs doublet scalar before electroweak symmetry breaking:
\begin{equation}\label{eq:d6}
V_\mathrm{self} = \mu^2 |H|^2 + \lambda |H|^4 + \mathcal{O}_6,\; \mathcal{O}_6 \equiv \frac{c_6}{\Lambda^2}
  \lambda |H|^6,
\end{equation}
where $\mu^2$ and $\lambda$ are the conventional parameters employed
in the SM potential for the Higgs doublet $H$.

The changes in the quartic and the triple Higgs couplings, defined in Eq.~\ref{eq:indep}, are
related via~\cite{Goertz:2014qta}:\footnote{Note that, in general, $c_3$ and $d_4$
  would be multiplied by $v^2 / \Lambda^2$ in $D=6$ EFT. We have set
  $\Lambda = v$ for simplicity here.}
\begin{align}\label{eq:selfcoup}
c_3 = c_6 ,\; d_4 =  6 c_6 \, .
\end{align}

Due to the relation appearing in Eq.~\ref{eq:selfcoup}, the cross section for triple Higgs boson production is a quartic polynomial in $c_6$, i.e. it contains terms up to $c_6^4$. Such terms come from squared matrix elements of diagrams containing two triple Higgs couplings, such as the one shown in Fig.~\ref{fig:feyndiags}(d). 

\begin{figure*}[!htb]
  \centering
   \subfigure[]{\includegraphics[width=0.49\textwidth]{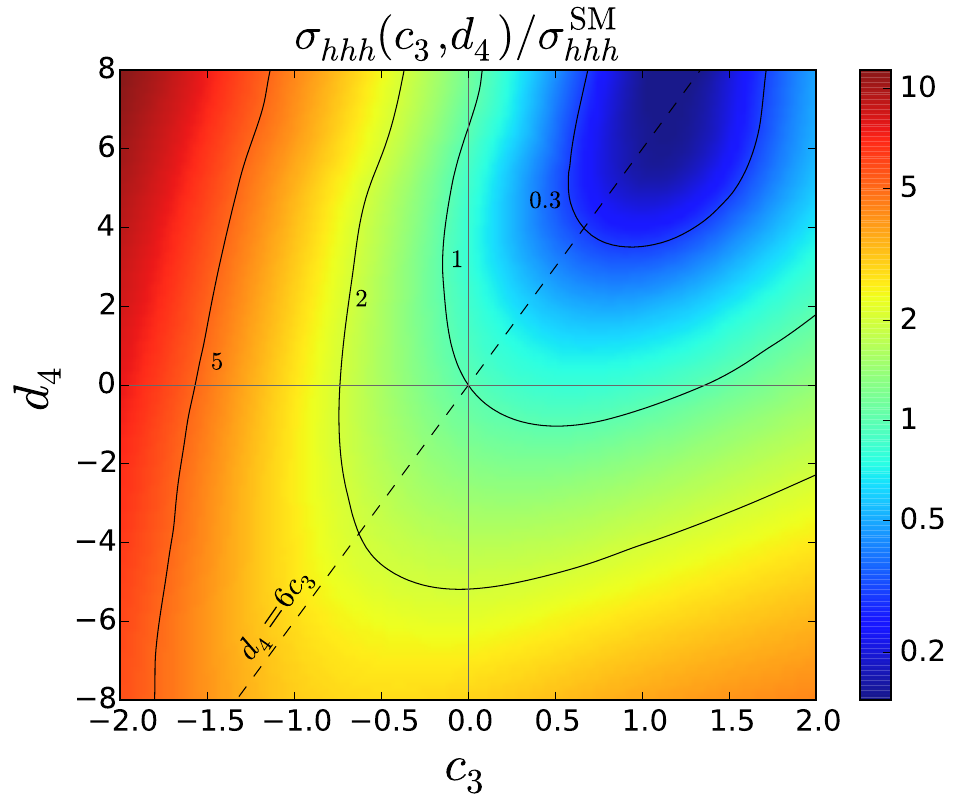}}
   \hfill
   \subfigure[]{\includegraphics[width=0.49\textwidth]{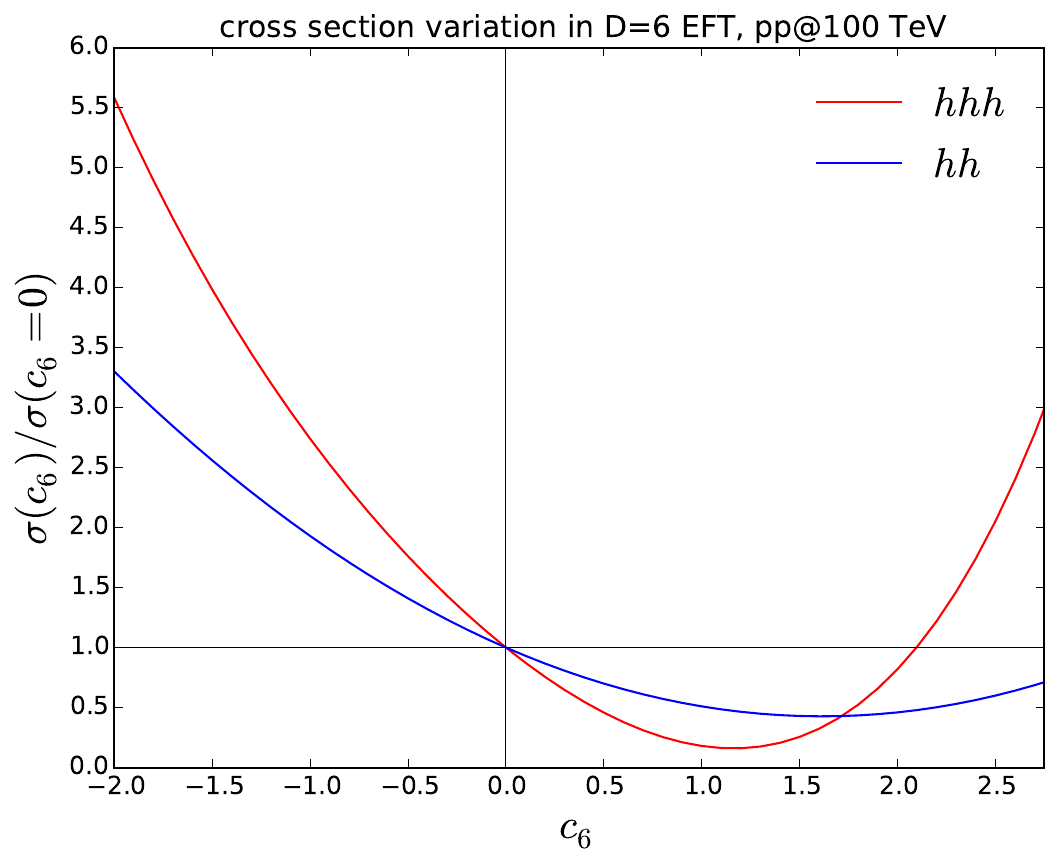}}
  \caption{\label{fig:xsecs} Total cross section ratios normalised to
    the Standard Model values for gluon-fusion-initiated multi-Higgs production at
    100~TeV. The Higgs boson mass was fixed to $m_h = 125$~GeV. The SM
    cross section at leading order is $\sim 2.88$~fb. On the left-hand
    panel we show a contour plot of the variation of the cross section
    ratio with respect to the $c_3$ and $d_4$ parameters (see Eq.~\ref{eq:indep})). On the right-hand panel one can see the variation with respect to
the SM in a theory where the SM is extended with a $\mathcal{O}_6 \sim
|H|^6$ operator as in Eq.~\ref{eq:d6}, for both Higgs boson pair production ($hh$) and Higgs boson triple production ($hhh$). For both calculations, the
\texttt{NNPDF23\_nlo\_as\_0119} parton density function set was used.}
\end{figure*}

In Fig.~\ref{fig:xsecs}(b) we show the
variation of the inclusive leading-order cross sections for ggF $hh$ and $hhh$ with respect to
the SM ($c_6 = 0$).  The fit as a function of $c_6$ for
the two cases, at 100~TeV, is:
\begin{eqnarray}
\frac{\sigma(c_6)_{hh}}{\sigma(\mathrm{SM})_{hh}} &=& 0.22 \times c_6^2\nonumber\\
&-& 0.71\times c_6 + 1.00, \nonumber \\
\frac{\sigma(c_6)_{hhh}}{\sigma(\mathrm{SM})_{hhh}}  &=&  0.03 \times c_6^4\nonumber \\
&+&  0.03 \times c_6^3 + 0.43 \times c_6^2  \nonumber\\
&-&1.31 \times c_6 + 1.00.
\end{eqnarray}

The line $d_4 = 6c_3$ is also shown as a dissection on the $c_3 - d_4$ plane in Fig.~\ref{fig:xsecs}(a). 

\section{Triple Higgs production final states}
\label{sec:fs}

We list the dominant Higgs boson triple production final states, i.e. those that yield $N_{\rm events} > $ 10 with
30\,ab$^{-1}$ of integrated luminosity at a proton collider at 100~TeV
centre-of-mass energy, in Table~\ref{tab:channel}.

\begin{table}[!htp]
\begin{ruledtabular}
\begin{tabular}{llll}
       $hhh \to $ final state  &  BR (\%) &   $\sigma$ (ab) & $N_{30 \mathrm{ab}^{-1}}$  \\\hline
                       $ (b \bar b) (b \bar b) (b \bar b) $  &  19.21  &    1110.338 &      33310  \\
                    $ (b \bar b) (b \bar b) (WW_{1 \ell}) $  &  7.204  &      416.41 &      12492  \\
                 $ (b \bar b) (b \bar b) (\tau \bar \tau) $  &  6.312  &     364.853 &      10945  \\
              $ (b \bar b) (\tau \bar \tau) (WW_{1 \ell}) $  &  1.578  &       91.22 &       2736  \\
                    $ (b \bar b) (b \bar b) (WW_{2 \ell}) $  &  0.976  &      56.417 &       1692  \\
                 $ (b \bar b) (WW_{1 \ell}) (WW_{1 \ell}) $  &  0.901  &      52.055 &       1561  \\
           $ (b \bar b) (\tau \bar \tau) (\tau \bar \tau) $  &  0.691  &      39.963 &       1198  \\
                    $ (b \bar b) (b \bar b) (ZZ_{2 \ell}) $  &  0.331  &      19.131 &        573  \\
                 $ (b \bar b) (WW_{2 \ell}) (WW_{1 \ell}) $  &  0.244  &      14.105 &        423  \\
                  $ (b \bar b) (b \bar b) (\gamma \gamma) $  &  0.228  &      13.162 &        394  \\
              $ (b \bar b) (\tau \bar \tau) (WW_{2 \ell}) $  &  0.214  &      12.359 &        370  \\
           $ (\tau \bar \tau) (WW_{1 \ell}) (WW_{1 \ell}) $  &  0.099  &       5.702 &        171  \\
        $ (\tau \bar \tau) (\tau \bar \tau) (WW_{1 \ell}) $  &  0.086  &       4.996 &        149  \\
                 $ (b \bar b) (ZZ_{2 \ell}) (WW_{1 \ell}) $  &  0.083  &       4.783 &        143  \\
              $ (b \bar b) (\tau \bar \tau) (ZZ_{2 \ell}) $  &  0.073  &       4.191 &        125  \\
               $ (b \bar b) (\gamma \gamma) (WW_{1 \ell}) $  &  0.057  &       3.291 &         98  \\
            $ (b \bar b) (\tau \bar \tau) (\gamma \gamma) $  &   0.05  &       2.883 &         86  \\
              $ (WW_{1 \ell}) (WW_{1 \ell}) (WW_{1 \ell}) $  &  0.038  &       2.169 &         65  \\
           $ (\tau \bar \tau) (WW_{2 \ell}) (WW_{1 \ell}) $  &  0.027  &       1.545 &         46  \\
     $ (\tau \bar \tau) (\tau \bar \tau) (\tau \bar \tau) $  &  0.025  &       1.459 &         43  \\
                 $ (b \bar b) (WW_{2 \ell}) (WW_{2 \ell}) $  &  0.017  &       0.956 &         28  \\
              $ (WW_{2 \ell}) (WW_{1 \ell}) (WW_{1 \ell}) $  &  0.015  &       0.882 &         26  \\
                    $ (b \bar b) (b \bar b) (ZZ_{4 \ell}) $  &  0.012  &        0.69 &         20  \\
        $ (\tau \bar \tau) (\tau \bar \tau) (WW_{2 \ell}) $  &  0.012  &       0.677 &         20  \\
                 $ (b \bar b) (ZZ_{2 \ell}) (WW_{2 \ell}) $  &  0.011  &       0.648 &         19  \\
           $ (\tau \bar \tau) (ZZ_{2 \ell}) (WW_{1 \ell}) $  &  0.009  &       0.524 &         15  \\
               $ (b \bar b) (\gamma \gamma) (WW_{2 \ell}) $  &  0.008  &       0.446 &         13  \\
         $ (\tau \bar \tau) (\gamma \gamma) (WW_{1 \ell}) $  &  0.006  &        0.36 &         10  \\
\end{tabular}
\caption{The list of channels with $N_{\rm events} > $ 10 with
 30\,ab$^{-1}$ and their branching ratios (BR). The subscript ``$_{x\ell}$'' denotes the number of
 leptons $x$ in the final state, originating from the di-bosons. The
 cross section used for $pp\rightarrow hh$ at 100~TeV is
 $\sigma_\mathrm{NLO} = \sigma_\mathrm{LO} \times 2.0 =
 5.78$~fb, where a $K$-factor $K=2.0$ has been applied to obtain an estimate of the NLO cross section. The number of events has been rounded to the nearest
 integer.}
\label{tab:channel}
\end{ruledtabular}
\end{table}%

If we apply further requirements to the final states listed in
Table~\ref{tab:channel}:
\begin{itemize}
\item to possess greater than 100 events at 30~ab$^{-1}$ of integrated
  luminosity,
\item and all gauge bosons fully decay to leptons,
\end{itemize}
then we are left with the following interesting final states:
$(b\bar{b})(b\bar{b}) (b\bar{b})$, $(b\bar{b})(b\bar{b}) (\tau
\bar{\tau})$, $(b\bar{b})(b\bar{b}) (WW_{2\ell})$, $(b\bar{b}) (\tau
\bar{\tau}) (\tau\bar{\tau})$, $(b\bar{b})(b\bar{b}) (\gamma \gamma)$,
$(b\bar{b})(\tau \bar{\tau}) (WW_{2\ell})$. In particular, the expected
combined number of events in the multi-$b$-jet and multi-$\tau$ final
states is $\sim$45000 over the lifetime of the FCC-hh, and will most
likely provide valuable information on the triple Higgs boson process. In the present study we focus on the rare but clean final state
$(b\bar{b})(b\bar{b}) (\gamma \gamma)$. 

\section{Event generation and detector simulation}
\label{sec:sim}
\subsection{Detector simulation}
In the hadron-level analysis that follows, we consider all particles within a pseudorapidity of
$|\eta| < 5$ and $p_T > 400$~MeV. We reconstruct jets using the anti-$k_t$ algorithm available in the
\texttt{FastJet} package~\cite{Cacciari:2011ma, Cacciari:2005hq}, with a radius
parameter of $R=0.4$. We only consider jets with $p_T > 40$~GeV within
$|\eta| < 3.0$ in our analysis. We consider photons within $|\eta| < 3.5$
and $p_T > 40$~GeV and 100\% reconstruction efficiency. The jet-to-photon mis-identification probability is
taken to be $\mathcal{P}_{j\rightarrow \gamma} = 10^{-3}$, flat over all
momenta above the $p_T$ cut and over all pseudorapidities.\footnote{Note that the HL-LHC expectation has the
  approximate form $\mathcal{P}_{j\rightarrow \gamma} = 0.0093 \times
 \mathrm{e}^{-0.036 p_{Tj}/\mathrm{GeV}}$~\cite{Barr:2014sga}. For a
 $p_T \sim 40$~GeV, this gives approximately
 $\mathcal{P}_{j\rightarrow \gamma}  \sim 2\times 10^{-3}$. Thus, the
 value employed here is expected to be a reasonable approximation to
 future detector performance.} We also consider the mis-tagging of two light jets to bottom-quark-initiated jets with a flat probability of 1\% for each mis-tag, corresponding to a flat $b$-jet
identification rate of 80\% and demand that they lie within $|\eta| <
3.0$. We demand all photons to be isolated, an isolated photon having $\sum_i p_{T,i}$
less than 15\% of its transverse momentum in a cone of $\Delta R =
0.2$ around it. Finally, no detector-smearing effects have been considered. 

\subsection{Event generation}
Events for the $hhh$ signal samples have been generated via the loop-induced
module of the \texttt{MadGraph 5/aMC@NLO} package~\cite{Frixione:2010ra,
  Frederix:2011zi, Alwall:2014hca, Alwall:2014bza,
  Hirschi:2015iia}. The SM loop model present in \texttt{MadGraph 5/aMC@NLO} was modified to allow for
deformations of the Higgs boson triple and quartic self-couplings away from the SM values. All tree-level and next-to-leading order (i.e. matched via the MC@NLO method~\cite{Frixione:2002ik}) background processes have been generated
using \texttt{MadGraph 5/aMC@NLO}, apart from the di-Higgs plus jets
($hh$ + jets) background, which was simulated using \texttt{HERWIG++} in conjunction with the \texttt{OpenLoops}
matrix-element generator~\cite{Cascioli:2011va, Maierhofer:2013sha}. The default parton density
functions were used in each case: for the signal and tree-level
backgrounds (including $hh$+jets) the \texttt{NNPDF23\_nlo\_as\_0119} set was used, whereas for
the NLO samples the \texttt{NNPDF23\_nlo\_as\_0118\_qed} set was
employed~\cite{Ball:2012cx}.

Due to the large cross sections and high-multiplicity final states present at a
100~TeV collider, we only generate the tree-level processes to include true photons and true $b$-quarks at
parton level. This implies that light extra jets for these processes will be generated by
the parton shower, for which we employ the \texttt{HERWIG++} general-purpose
event generator~\cite{Bahr:2008pv, Gieseke:2011na, Arnold:2012fq, Bellm:2013lba}.\footnote{Simulation of hadronization and the underlying event
were also included.~\cite{Bahr:2008dy}. No simulation of pile-up
events was considered.} Inevitably this
introduces an uncertainty to the results presented herein, rendering
any observables related to these light jets leading-log accurate.\footnote{The $hh$+jets process is the only exception, with the first jet being leading-order accurate~\cite{Maierhofer:2013sha}.} We do not
expect this, however, to alter the main conclusions of this first, baseline, study. Furthermore, generation-level cuts that anticipate the analysis
cuts at hadron level are imposed on the $b$ quarks and the photons. In
the case of decaying resonances (i.e. $h$ and $Z$ bosons) no cuts are
imposed. The phase-space cuts applied on the samples
$b\bar{b} b\bar{b}$, $b\bar{b} b\bar{b} \gamma$, $b\bar{b}
b\bar{b}\gamma\gamma$, $b\bar{b} \gamma\gamma$ are shown in Table~\ref{tab:pscuts}.
\begin{table}[!htp]
\begin{ruledtabular}
\begin{tabular}{ll}
 observable & PS cut \\\hline
 $p_{T,b}$& $>35~\mathrm{GeV}$, at least one $> 70~\mathrm{GeV}$\\
 $|\eta _b|$ & $< 3.2$\\ 
$p_{T,\gamma}$ & $>35~\mathrm{GeV}$, at least one $> 70~\mathrm{GeV}$\\
$|\eta _\gamma|$ & $<3.5$\\
$\Delta R_{\gamma\gamma}$ & $>0.2$  \\
$m_{\gamma \gamma}$ & $\in [90, 160]~\mathrm{GeV}$\\
\end{tabular}
\caption{The phase-space (PS) cuts imposed on the background samples $b\bar{b} b\bar{b}$, $b\bar{b} b\bar{b} \gamma$, $b\bar{b}
b\bar{b}\gamma\gamma$, $b\bar{b} \gamma\gamma$.}
\label{tab:pscuts}
\end{ruledtabular}
\end{table}%


At this point one should stress that even though NLO event generation
matched to the parton shower has been largely automated, NLO calculations for the high-multiplicity final
states, particularly with many coloured particles and complicated
phase space cuts, remain challenging at
present. We hence apply a conservatively large flat $K$-factor of
$K=2.0$ to all the processes calculated at tree level, as well as the
$hhh$ and $hh$+jets loop-induced processes. This is a crucial point that should be addressed in future studies at higher-energy hadron colliders, as
such final states will become increasingly common. 

The analysis of the signal and backgrounds generated for the final state $ (b \bar{b})
(b \bar{b}) (\gamma\gamma)$ is presented in section~\ref{sec:analysis1}.

\section{Analysis}
\label{sec:anal}
\subsection{Differential distributions}
We investigate the shape of the differential distributions in Higgs
triple production in the Standard Model. Here we keep the Higgs bosons stable and include
parton shower effects. We compare the shape of the $hhh$ distributions
to those coming from the more familiar case of Higgs boson pair production ($hh$) at
100~TeV. 

\begin{figure*}[!htp]
  \centering
   \subfigure[]{\includegraphics[width=0.45\textwidth]{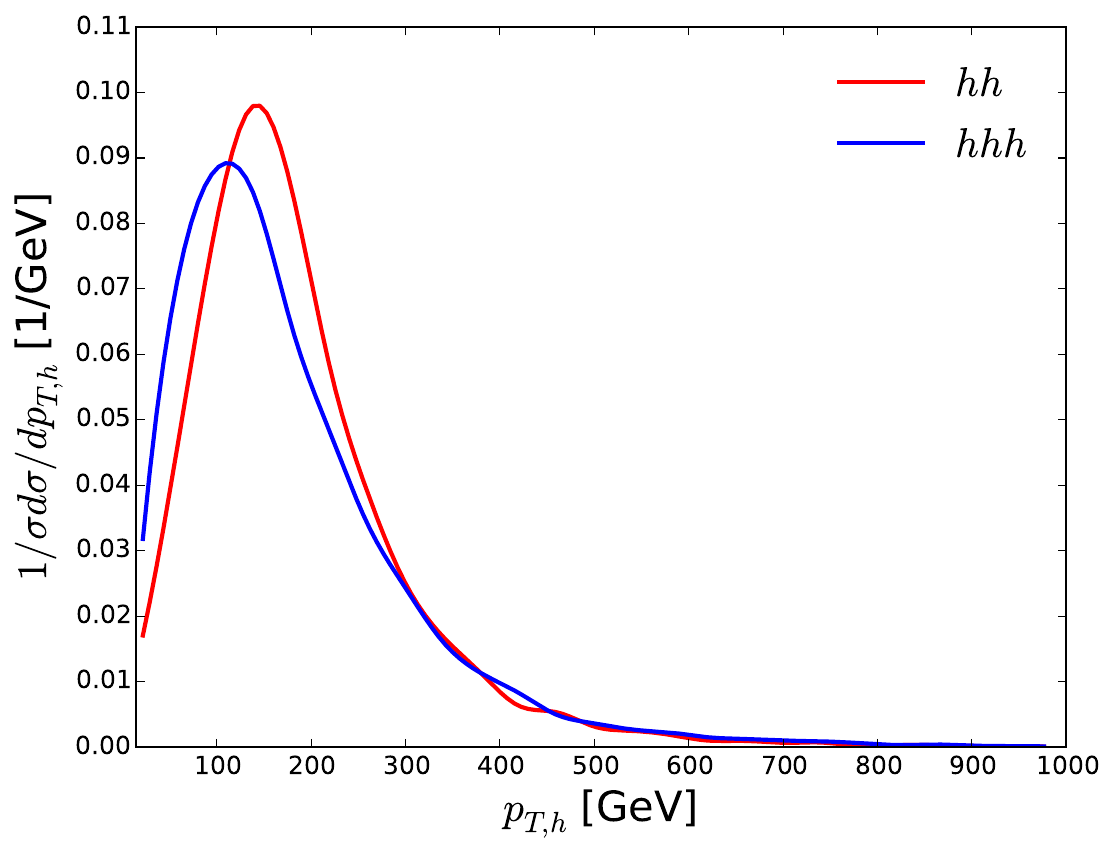}}
  \hfill
   \subfigure[]{\includegraphics[width=0.45\textwidth]{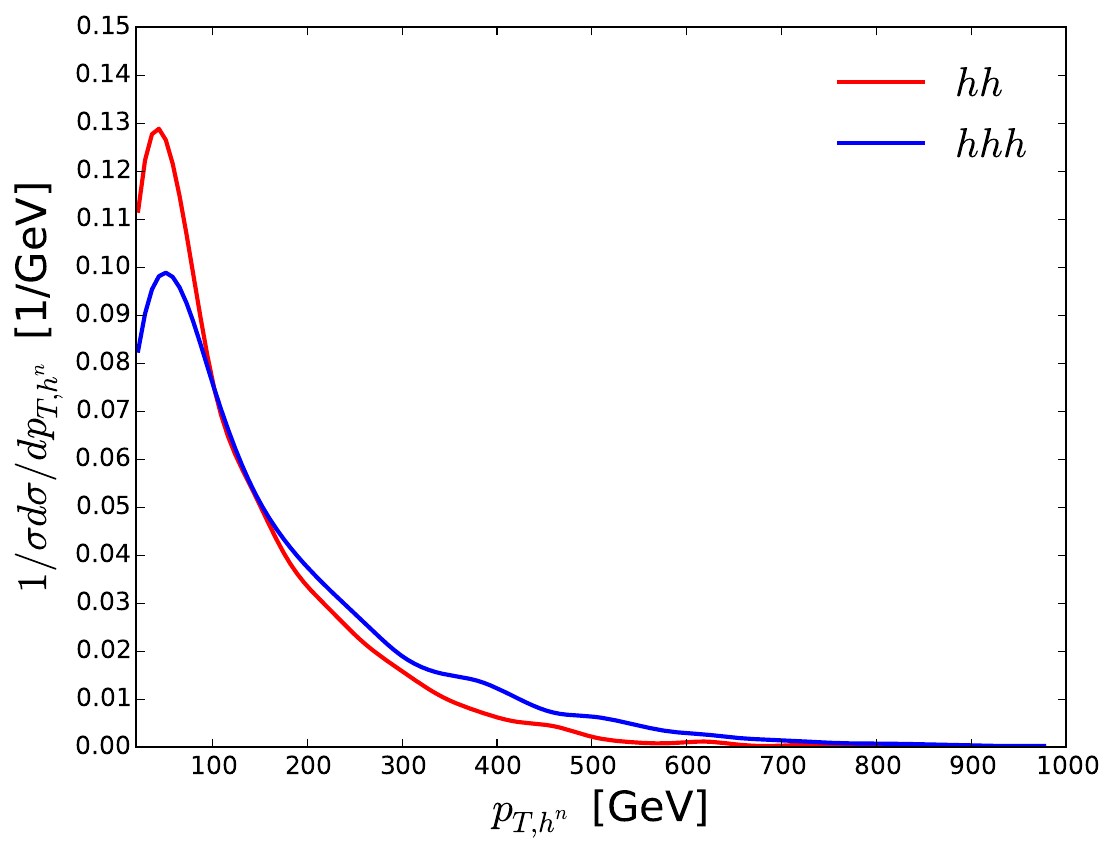}}
  \hfill\\
   \subfigure[]{\includegraphics[width=0.45\textwidth]{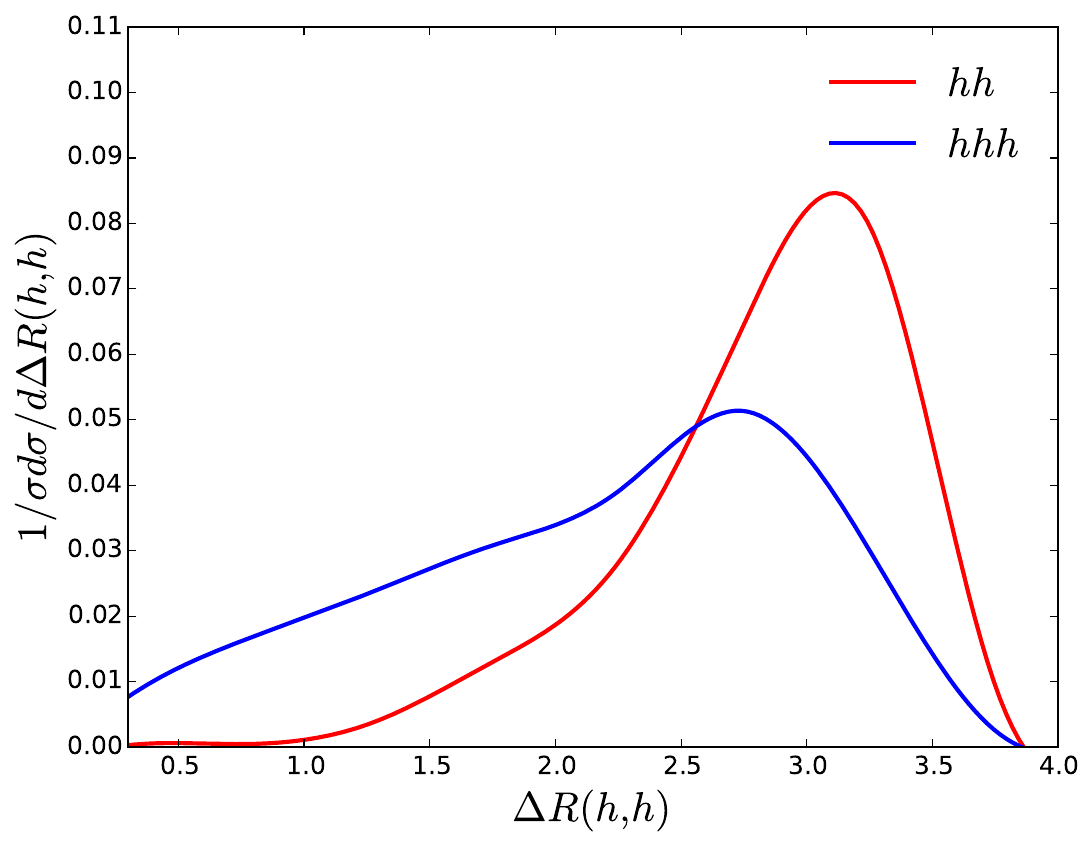}}
  \hfill
   \subfigure[]{\includegraphics[width=0.45\textwidth]{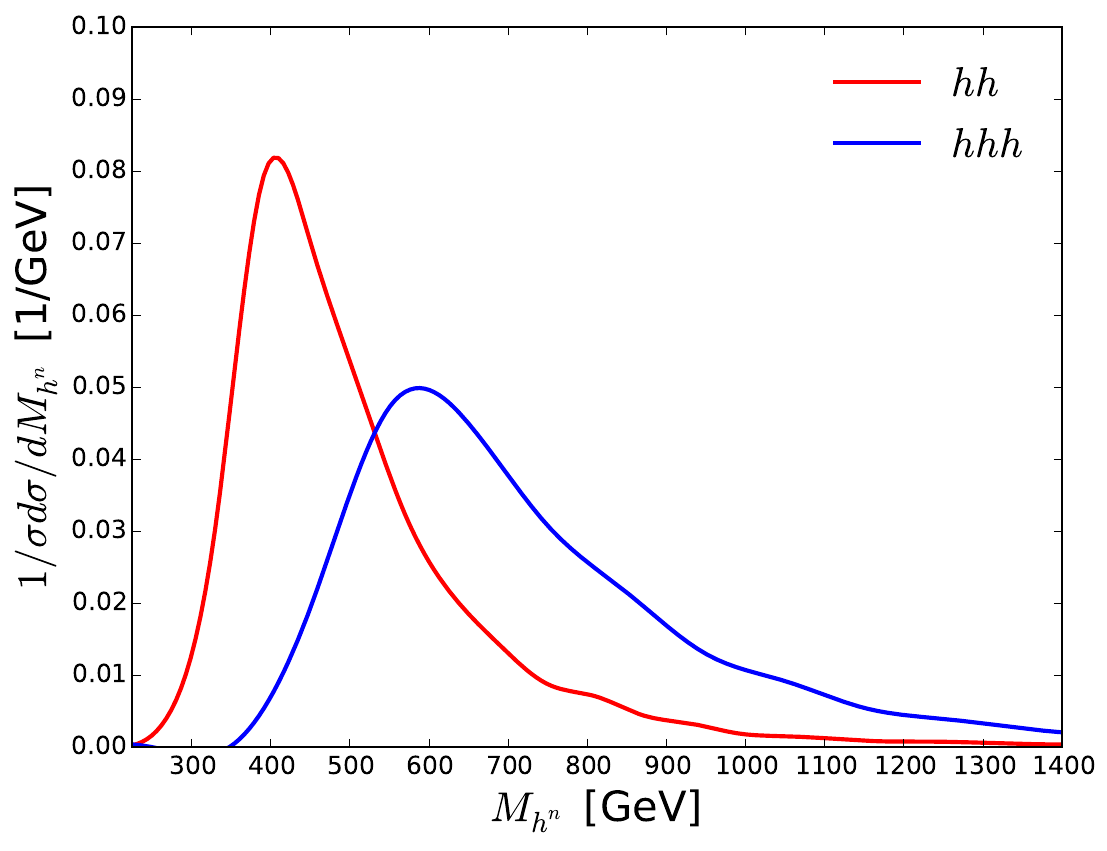}}
  \caption{\label{fig:comparison} Comparison of differential
    distributions for Higgs boson pair ($hh$) and triple production
    ($hhh$) in the Standard Model. Parton showering effects are
    included on top of leading-order matrix elements. Figure~(a) shows the transverse momentum of any
single Higgs boson, $p_{T,h}$. In (b) we show the the spectrum of the
transverse momentum of the Higgs boson ``system'', $p_{T,h^n}$, i.e. the triplet of
Higgs bosons in $hhh$, and the two Higgs bosons in $hh$. In (c) the distance between two Higgs bosons, $\Delta R (h,h)$, is examined and in (d) we show the the invariant mass of all Higgs
bosons, $M_{h^n}$.}
\end{figure*}

Figure~\ref{fig:comparison}(a) shows the transverse momentum of any
single Higgs boson either in $hh$ or $hhh$ production, $p_{T,h}$. Evidently, the transverse momentum of a Higgs boson in $hhh$ is softer than that
of $hh$, peaking at $\sim 100$~GeV instead of $\sim 150$~GeV. 

In Fig.~\ref{fig:comparison}(b) we show the the spectrum of the
transverse momentum of the Higgs boson ``system'', $p_{T,h^n}$, i.e. the triplet of
Higgs bosons in $hhh$, and the two Higgs bosons in $hh$. One can
observe that the $p_{T,h^n}$ is harder in $hhh$ than that of the pair in
$hh$. 

We examine the distance between two Higgs bosons, $\Delta R (h,h)$, in $hh$ and $hhh$
production in Fig.~\ref{fig:comparison}(c). In the case of triple production the distance is calculated between any
two Higgs bosons. The Higgs bosons in $hh$ are found to be more back-to-back than
those in $hhh$, as expected. 

Finally, in Fig.~\ref{fig:comparison}(d) we show the the invariant mass of all Higgs
bosons in $hh$ or $hhh$ production, $M_{h^n}$. The invariant mass
distribution in $hhh$ peaks just above $M_{h^3} \sim 600$~GeV, whereas
that in Higgs pair production, just above $M_{h^2} \sim 400$~GeV. 

\subsection{$hhh \to (b \bar{b})  (b \bar{b}) (\gamma\gamma)$}
\label{sec:analysis1}
The $hhh \to (b \bar{b})  (b \bar{b}) (\gamma\gamma)$ process is
expected to be relatively clean and simple to
reconstruct.\footnote{Note that this final state has been considered
  in~\cite{Barger:2014qva}, in the context of the two-Higgs doublet model $hH
  \rightarrow hhh$ final state. Here we consider the SM case.} The
excellent resolution of the di-photon invariant mass, that has
contributed to the Higgs boson discovery at the LHC's Run 1, can be
exploited to facilitate background rejection. 

The present analysis follows a simple path, using the $R=0.4$ anti-$k_t$ jets
as described in Section~\ref{sec:sim}. Note, however, that an analysis utilising the jet substructure of boosted Higgses to a bottom-anti-bottom pairs, e.g. as in~\cite{deLima:2014dta}, could assist in
signal-background separation. We defer this task to future work. 

We ask for four $b$-jets, or light jets mis-identified as $b$-jets,
within $|\eta | < 3.0$, possessing transverse momenta $p_{T,b _{\{1,2,3,4\}}} > \{ 80, 50, 40, 40
\}$~GeV, where the subscripts 1, 2, 3, 4 denote the first, second,
third and fourth hardest $b$-jets respectively. We ask for two photons, or
mis-identified jets as photons, within $|\eta| < 3.0$ and
$p_{T,\gamma_{\{1,2\}}} > \{ 70, 40\}$~GeV. Due to the fact that, for
the majority of $b$-jets we cannot identify whether they originated
from a $b$-quark or an anti-$b$-quark, there exists a $3$-fold combinatorial
ambiguity in combining $b$-jets into the two Higgs boson candidates. As a simple
choice, we take the highest-$p_T$ $b$-jet and pair it with the closest
$b$-jet in $\Delta R = \sqrt{ \Delta \eta^2 + \Delta \phi^2} $, and
pair the other two remaining $b$-jets together.\footnote{We have
  verified explicitly that an alternative method based on minimization
  of the squared sum of $(m_{bb}-m_h)$ from each combination yields results that differ by
  $\mathcal{O}(1\%)$ compared to the simpler $\Delta R$ method.} We thus
construct the paired $b$-jet invariant mass, respectively, $m_{bb}^{\mathrm{close},1}$ and
$m_{bb}^{\mathrm{close},2}$, for which we demand
$m_{bb}^{\mathrm{close},1} \in [100, 160]$~GeV and
$m_{bb}^{\mathrm{close},2} \in [90, 170]$~GeV. The rather large mass
windows are chosen to maintain high signal efficiency given the small
initial cross section. Moreover, we construct
the distance between the highest-$p_T$ $b$-jet and the corresponding
paired one, and impose $\Delta R_{bb}^{\mathrm{close}, 1} \in [ 0.2,
1.6 ]$.\footnote{The distance between the other paired $b$-jets was
  not found to have significant discriminatory power.} For the photon pair, we simply construct the invariant mass and impose a strong window on the measured Higgs boson mass $m_{\gamma\gamma} \in [ 124, 126 ]
$~GeV.\footnote{This cut implies that the di-photon resolution should be better
than $\sim 1$~GeV at the FCC-hh. The current resolution at the LHC is
1-2~GeV,~\cite{Chatrchyan:2012twa, Aad:2014eha} and thus it is not
unreasonable to expect an improvement at the detectors of the
future collider.} We also restrict the distance between the two photons to $\Delta R_{\gamma
  \gamma} \in [ 0.2, 4.0 ]$. We collect these selection cuts in Table~\ref{tab:cuts}.

\begin{table}[!htp]
\begin{ruledtabular}
\begin{tabular}{ll}
 observable & selection cut \\\hline
 $p_{T,b _{\{1,2,3,4\}}}$ & $> \{ 80, 50, 40, 40\}$~GeV\\
 $|\eta _b|$ & $< 3.0$\\ 
$m_{bb}^{\mathrm{close},1}$ & $\in [100, 160]$~GeV \\
$m_{bb}^{\mathrm{close},2}$ & $\in [90, 170]$~GeV \\
$\Delta R_{bb}^{\mathrm{close}, 1}$ &  $\in [ 0.2, 1.6 ]$ \\
$\Delta R_{bb}^{\mathrm{close}, 2}$ &  no cut\\
$p_{T,\gamma_{\{1,2\}}}$ & $> \{ 70, 40\}$~GeV\\
$|\eta _\gamma|$ & $<3.5$\\
$\Delta R_{\gamma\gamma}$ & $\in [ 0.2, 4.0 ]$  \\
$m_{\gamma \gamma}$ & $\in [124, 126]~\mathrm{GeV}$\\
\end{tabular}
\caption{The final selection cuts imposed in the analysis of the $(b \bar{b})  (b \bar{b}) (\gamma\gamma)$ final state. The observables are defined in the main text. }
\label{tab:cuts}
\end{ruledtabular}
\end{table}%

\begin{table*}[!htp]
\begin{ruledtabular}
\resizebox{\linewidth}{!}{
\begin{tabular}{lllll}
process & $\sigma_\mathrm{LO}$ (fb) & $\sigma_\mathrm{NLO}  \times \mathrm{BR} \times \mathcal{P}_\mathrm{tag} $ (ab) & $\epsilon_\mathrm{analysis}$ & $N^{\mathrm{cuts}}_{30~\mathrm{ab}^{-1}}$  \\\hline
$hhh \to (b \bar{b})  (b \bar{b}) (\gamma\gamma)$, SM &  2.89  &  5.4 & 0.06 & 9.7\\
$hhh \to (b \bar{b})  (b \bar{b}) (\gamma\gamma)$, $c_6 = 1.0$ &  0.46  & 0.9 & 0.04 & 1.1 \\
$hhh \to (b \bar{b})  (b \bar{b}) (\gamma\gamma)$, $c_6 = -1.0$ &  7.94  & 15.0  &  0.05 & 22.5 \\\hline
$b \bar{b} b \bar{b} \gamma\gamma$ &  1.28  &  1050 &  $2.6\times10^{-4}$ & 8.2\\
$hZZ$, (NLO) $(ZZ\rightarrow (b\bar{b})(b\bar{b}))$ &  0.817  & 0.8 & 0.002 & $\ll 1$\\
$hhZ$, (NLO)$(Z \rightarrow (b\bar{b}))$  &  0.754 & 0.8 & 0.007  & $\ll1$\\
$hZ$, (NLO) $(Z \rightarrow (b\bar{b}))$ & $8.019\times 10^3$& 1129& $\mathcal{O}(10^{-5})$& $\ll 1$\\ 
$b \bar{b} b \bar{b} \gamma$ + jets &  $2.948 \times 10^3$  &  2420 & $\mathcal{O}(10^{-5})$ &$\mathcal{O}(1)$\\
$b \bar{b} b \bar{b}$ + jets &  $5.449 \times 10^3$ & 4460 & $\mathcal{O}(10^{-6})$& $\ll 1$\\
$b \bar{b} \gamma\gamma$ + jets &  98.7  &  4.0 & $\mathcal{O}(10^{-5})$ & $\ll 1$ \\\hline
$hh$ + jets, SM & 275.0 &  592.7 & $7\times 10^{-4}$  & 12.4 \\
$hh$ + jets , $c_6 = 1.0$ & 153.8  &  331.5 & 0.001 & 9.9 \\
$hh$ + jets , $c_6 = -1.0$ & 518.2 &  1116.9& $4\times 10^{-4}$ & 13.4 \\
\end{tabular}
}
\end{ruledtabular}
\caption{The processes considered in the analysis of the $(b \bar{b})
  (b \bar{b}) (\gamma\gamma)$ final state. The parton-level cross
  section, including the cuts given in the main text is given (if
  any), the analysis efficiency and the
  expected number of events at $30$~ab$^{-1}$ are given. A flat
  $K$-factor of $K=2.0$ has been applied to all tree-level processes (including
  $hh$+jets) as an estimate of the expected increase in cross section
  from LO to NLO. The $hZZ$, $hhZ$ and $hZ$ processes have been produced at
  NLO and hence no $K$-factor is applied. Even though the $hhZ$ process depends on $c_6$, we only consider the SM case, as it was found to be negligible after cuts.}
\label{tab:procs}
\end{table*}

We show a summary of the processes considered in the analysis in Table~\ref{tab:procs}. The most significant backgrounds in our set-up turn out to be the SM $b \bar{b} b \bar{b} \gamma\gamma$ and those coming from Higgs boson pair production in association with extra jets. Specifically, the latter emulates the signal well, as the di-photon mass window is expected to have similar efficiency to the signal. Moreover, as we have pointed out at the beginning of the section, the Higgs bosons in $hh$ are \textit{harder} on average than than those in $hhh$, thus passing transverse momentum cuts easily. This background could be tackled in future studies via $h\rightarrow b\bar{b}$ tagging using jet substructure techniques that exploit the decay versus the $g \rightarrow b\bar{b}$ branching that produces the additional $b\bar{b}$ pair in $hh$+jets.\footnote{Note that the additional two $b$-jets in $hh$+jets and $hZ$ have been generated by gluon splitting into $b\bar{b}$, performed by the shower Monte Carlo.}

\subsection{Sensitivity in $D=6$ EFT}\label{sec:d6}
Despite the rather large backgrounds, a signal-to-background ratio of $\mathcal{O}(1)$ can be obtained for the SM case. To summarise the results of the analysis, we present in the first two columns of Table~\ref{tab:results}, respectively, the number of expected $hhh$ events and  the total expected number of events, for the SM, as well as for the two simple deformations obtained by including the $D=6$ operator $\mathcal{O}_6$, with coefficient values $c_6 = \pm 1$. The third column of Table~\ref{tab:results} indicates that, if one assumes that the SM is the underlying theory, then $c_6 = \pm 1$ can be excluded at 95\% C.L. or better, using $hhh \rightarrow (b\bar{b}) (b\bar{b}) (\gamma\gamma)$ at the `high-luminosity' phase of the FCC-hh.

\begin{figure}[!htp]
  \centering
   \subfigure[]{\includegraphics[width=0.49\textwidth]{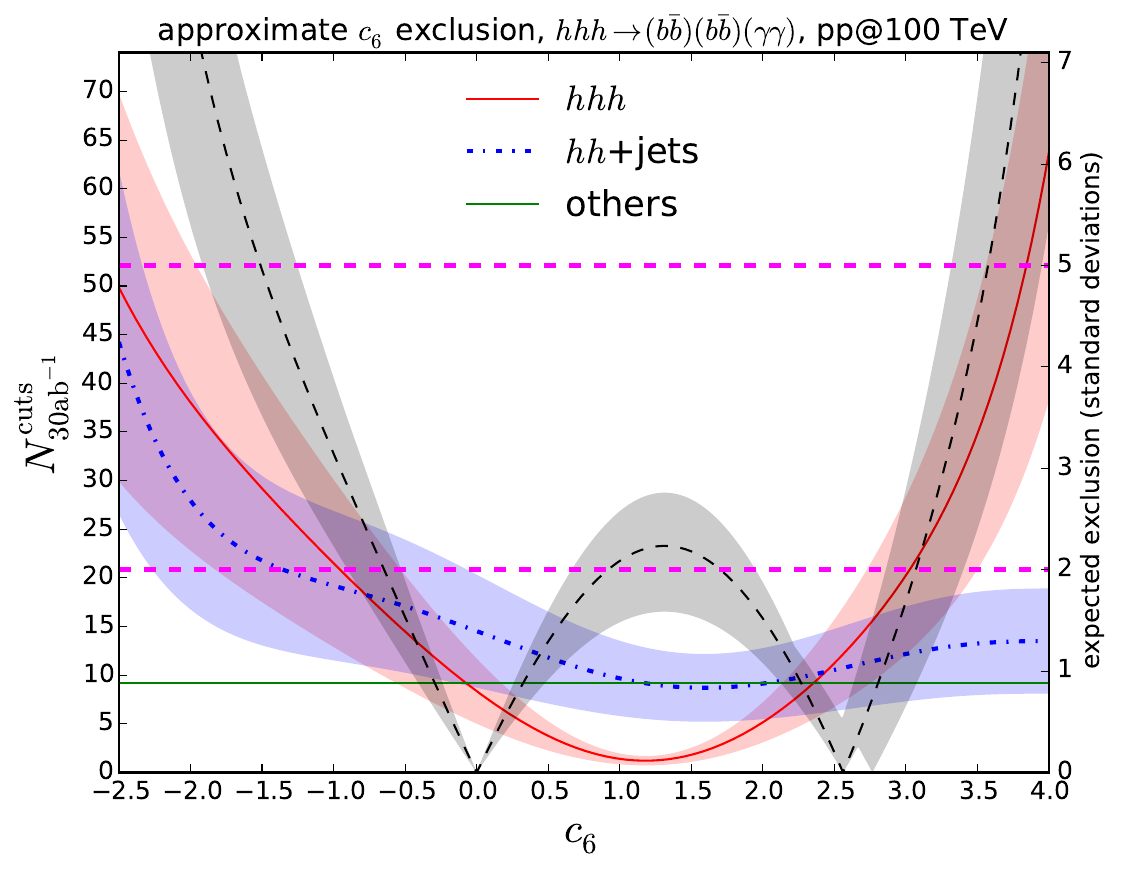}}
  \hfill
  \caption{\label{fig:projection} The expected exclusion significance
    on the $c_6$ coefficient (right vertical axis), assuming that the
    theoretical uncertainty on the expected number of $hhh$ and
    $hh$+jets events is $40\%$ for each process and uncorrelated
    between the two. The left vertical axis shows the expected number
    of events after cuts at 30~ab$^{-1}$. 
The horizontal magenta dashed lines show the $2\sigma$ and $5\sigma$ exclusion points. }
\end{figure}

Furthermore, we show in Fig.~\ref{fig:projection} the expected exclusion region on the $c_6$ coefficient, as well as the expected number of events after cuts, at 30~ab$^{-1}$. The theoretical uncertainty on the expected number of events for the $hh$ and the $hh$+jets processes was taken to be $40\%$ and uncorrelated between the two. The analysis efficiencies for $hhh$ and $hh$+jets were individually fitted using points in the region $c_6 \in [-3.0, 4.0]$.\footnote{The fitting uncertainty is not shown in Fig.~\ref{fig:projection}.} We assume that there is negligible uncertainty on the `other' backgrounds, which are taken to consist of the $b\bar{b}b\bar{b}\gamma\gamma$ and $b\bar{b}b\bar{b}\gamma$+jets processes. By examining the central values of the the grey exclusion band, we can see that the regions $c_6 \lesssim -0.7$ and $c_6 \gtrsim 3.0$, as well as the intermediate region $c_6 \in [\sim 1.0, \sim 1.7]$, are expected to be excluded at 95\% C.L. ($2\sigma$). Moreover, due to the fast-rising $hhh$ cross section, as a function of the $c_6$ coefficient in this simple model, the $5\sigma$-excluded region lies close to the $2\sigma$ outer regions: $c_6 \lesssim -1.4$, $c_6 \gtrsim 3.5$. Note that the analysis can be optimised for each value of $c_6$ to obtain a higher significance, but in light of the many sources of uncertainties we do not pursue this here. Such optimisation could substantially alter the shape of the $hhh$ and $hh$+jets curves in Fig.~\ref{fig:projection}. 

\begin{table}[!htp]
\begin{ruledtabular}
\begin{tabular}{lllll}
\
  & $hhh$ & total & $\frac{| N(\mathrm{SM}) - N(c_6) |} { \sqrt{N(\mathrm{SM})}}$ \\\hline
 SM & 9.7 & 31.3  & \\
 $c_6 = 1.0$ & 1.1 & 20.2 & $\sim 2.0$ \\
 $c_6 = -1.0$ & 22.5 & 45.1 & $\sim 2.5$\\
\end{tabular}
\caption{The number of events for an integrated luminosity of 30~ab$^{-1}$ at 100~TeV, for the Standard Model and the the two simple deformations with $\mathcal{O}_6$, with coefficient values $c_6 = \pm 1$. The first and second columns show, respectively, the number of events for the $hhh$ signal and the total expected number of events for all contributing processes: $hhh$, $hh$+jets, $b\bar{b}b\bar{b}\gamma\gamma$ (using 8.2 events) and $b\bar{b}b\bar{b}\gamma$+jets (using 1 event). The third column shows, approximately, the level (in number of standard deviations) at which the two hypotheses $c_6 = \pm 1$ can be excluded given that the standard model is the underlying theory. }
\label{tab:results}
\end{ruledtabular}
\end{table}%

\subsection{Sensitivity on the $(c_3-d_4)$-plane}\label{sec:c3d4}
\begin{figure}[!htp]
  \centering
   \subfigure[]{\includegraphics[width=0.49\textwidth]{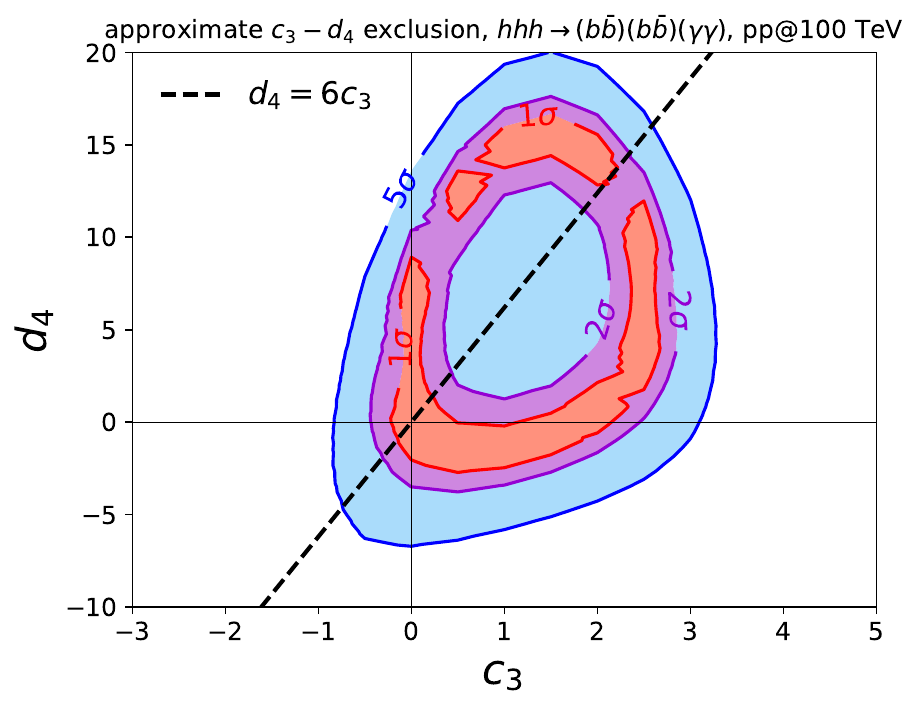}}
  \hfill
  \caption{\label{fig:projectionc3d4} The approximate expected
    $1\sigma$ (red), $2\sigma$ (violet) and $5\sigma$ (blue) exclusion regions
    on the $c_3-d_4$ plane after 30~ab$^{-1}$ of integrated luminosity, derived assuming a constant signal efficiency, calculated along the $d_4 = 6 c_3$ line in $c_3 \in [-3.0, 4.0]$.}
\end{figure}

Higgs boson triple production can be used to place constraints on the $(c_3-d_4)$-plane. This can subsequently be used to impose constraints on arbitrary relations between the triple and quartic coefficients in explicit models. We approximate the $hhh$ signal efficiency over the whole plane by calculating its average value for $c_3 \in [-3.0, 4.0]$, $d_4 = 6 c_3$, as obtained in the $D=6$ EFT example. The analysis is used verbatim, without any modification of cuts along the plane. The standard deviation on the efficiency obtained this way was found to be $\sim 20\%$ along this direction in the given interval. Considering the magnitude of the uncertainties on the signal and background predictions, we consider this to be adequate at present. For the $hh$+jets background we use the efficiency fit calculated for the $D=6$ EFT case. We show the projected constraints on the $(c_3-d_4)$-plane an integrated luminosity of 30~ab$^{-1}$ in Fig.~\ref{fig:projectionc3d4}. As a sanity check, we draw the $d_4 = 6 c_3$ line and check that the outer $2\sigma$-region: $c_6 \lesssim -2$ and $c_6 \gtrsim 3$ approximately reproduces the $D=6$ EFT result given the uncertainties. A few interesting observations can be made. Firstly, the whole region $c_3 \lesssim -1$ can be excluded at $5\sigma$ irrespective of the value of $d_4$ using triple Higgs production. Moreover, if $c_3$ is constrained to lie near $c_3 \sim 0$, then the weakest constraints on $d_4$ are obtained in all of the plane. On the other hand, if a non-zero value of $c_3$ is measured, e.g. $c_3 \sim 4$, then the constraint on $d_4$ can be quite stringent and in a region excluding $d_4 = 0$, i.e. $d_4 \in [\sim 4, \sim 8]$ at $5\sigma$.

\section{Discussion and Conclusions}
\label{sec:conc}
Evidently, discovering Standard Model-like triple Higgs boson production
will be a challenging task. Our analysis of the $hhh \to (b \bar{b})
(b \bar{b}) (\gamma\gamma)$ channel has demonstrated that the process merits serious investigation at a future collider running at 100 TeV proton-proton
centre-of-mass energy. It is important at this point to emphasise the
defining points and caveats that lead this phenomenological analysis
to this conclusion:
\begin{itemize}
\item The detector of an FCC-hh needs to have excellent photon
  identification and resolution, so that a di-photon invariant mass window of width
  2~GeV around the Higgs boson mass can imposed. As we already
  mentioned, the current resolution at the LHC is
1-2~GeV,~\cite{Chatrchyan:2012twa, Aad:2014eha}. Moreover, the projections for
  photon identification efficiency at the high-luminosity LHC are at
  $\mathcal{O}(80\%)$~\cite{TheATLAScollaboration:performance1}. It is
  not unreasonable to expect an improvement in both of these
  parameters at the FCC-hh, to a resolution of $\lesssim 1$~GeV or
  photon identification of $\gtrsim 90\%$.

\item Tagging of $b$-jets should be extremely good, at least in the range of
  70-80\%, with excellent light jet rejection of $\mathcal{O}(1\%)$
  over a wide range of transverse momenta and pseudorapidities. Reducing the
  tagging probability from 80\% to 70\% would reduce the final number
  of events in `true' 4-$b$-jet final states by about 40\%. We note
  that the expected performance of the $b$-tagging algorithms for the
  LHC Run 2 is already at this ballpark~\cite{ATL-PHYS-PUB-2015-022}.

\item Any analysis of triple Higgs production that includes $b\bar{b}$
  pairs will also benefit from a very good forward
  coverage, allowing identification of $b$-jets up to
  pseudo-rapidities of $|\eta| \sim 3.0$. Good forward coverage for photons to $|\eta| \sim 3.5$ would also benefit the analysis. For example, the fraction of signal events with two $b$-jets falling in $|
\eta_b| \in [2.5, 3.0]$ is $\sim 15\%$ and the fraction of events with two photons falling in $|\eta_\gamma| \in [2.5, 3.5]$ is $\sim 5\%$. These two are approximately uncorrelated, and thus an LHC-like coverage of  $|\eta_b| < 2.5$,  $|\eta_\gamma| < 2.5$ would cause a $\sim 20\%$ reduction in signal efficiency compared to the analysis presented in this article.  

\item Predictions of the triple Higgs boson production cross section, as for the case of double production, posses
  large theoretical uncertainties at present, due to the unknown
  higher-order corrections. The best available calculation includes only
  exact real emission diagrams in combination with `low-energy theorem'
  results~\cite{Maltoni:2014eza}. A full next-to-leading order
  calculation will reduce this and allow one to use the process to
  extract constraints on various models of new physics.

\item Crucially, the Monte Carlo event generation of multiple coloured partons (4-6) at
  next-to-leading order, with complicated phase-space cuts, matched to
  the parton shower, is essential. Technical
  improvements in this direction, along with increase in computing
  power, will allow us to perform predictions with reduced theoretical
  uncertainties, as well as perform analyses of more $hhh$ final states, such as those mentioned in Section~\ref{sec:fs} (as well as other processes that involve multiple Higgs bosons).

\item Due to the aforementioned theoretical and technical limitations, as well as the unknown characteristics of the future collider, we have not attempted to fully quantify the theoretical uncertainties permeating our results. We expect that future improvements in all of these aspects would allow one to obtain a more reliable quantitative result, including a reasonable expectation of uncertainty.

\end{itemize}

We note here that our event selection is optimised for the assumed
detector performance, and if some of these assumptions are changed,
the event selection should also be changed to optimise the signal
acceptance and background rejection. Moreover in the scenario that the FCC-hh performance is substantially worse
than what we have assumed, other channels could come into play, such
as $hhh \rightarrow (b\bar{b}) (b\bar{b}) (\tau^+ \tau^-)$ or $hhh
\rightarrow (b\bar{b}) (\tau^+ \tau^-)  (\tau^+ \tau^-)$. 

In conclusion, the study of triple Higgs production should be an important aspect of any
future collider programme. It could provide
complementary information on the nature of the Higgs boson and its
role in electroweak symmetry breaking, as well as extensions of the
Higgs boson sector beyond the standard model. This first baseline study resurrects this process and prompts further investigation into how it can be put into use.

\acknowledgments
We would like to thank Eleni Vryonidou, Paolo Torrielli and Valentin Hirschi for assistance with
Monte Carlo event generation as well as Jos\'e Zurita, Florian Goertz, Brian Batell and Jeremie Quevillon for providing useful comments and discussion. We would also like to thank the Physics
Institute, University of Z\"urich, for allowing continuous use of their computing
resources while this project was being completed. AP acknowledges support by the MCnetITN FP7 Marie Curie Initial Training Network
PITN-GA-2012-315877 and a Marie Curie Intra European Fellowship within the 7th European Community Framework Programme (grant no. PIEF-GA-2013-622071). KS is supported in part by the London Centre for Terauniverse Studies (LCTS), using funding from the European Research Council via the Advanced Investigator Grant 267352.


\bibliography{hhh.bib}
\bibliographystyle{apsrev4-1}

\end{document}